# Multifunctional Portable Optical Measuring Instrument Based on Y-Fiber Optics

Juntao He[1#], Yikai Dang[1#], Haoqi Wang[1#], Shaohua Wang[1#], Yingke Li[1#], Ruiyun Ma[1],

Yingyuan Li[1], Peilin Gao[1], Jianguo Cao[1], Yong Pan[1]*

1-College of Science, Xi'an University of Architecture and Technology, Xi'an 710055, China

*Corresponding author: panyong@xauat.edu.cn

**Abstract:** Based on grating diffraction principle, optical fiber transmission principle and optical interference principle, a multi-functional portable optical measuring instrument is constructed in this paper. The optical measurement visualization spectrometer based on CCD photoelectric image sensor is designed and assembled. The "Y" optical signal transmission fiber optical path suitable for multi-function measurement is improved and designed. The multi-function optical measurement system is built by combining with remote controlled multi-color LED lights. The spectral analysis, solution concentration monitoring and film thickness measurement are realized. The experimental results show that the observable wavelength range of the spectrometer is about 340-1050nm and the resolution is 1nm. The solution concentration can be obtained by measuring absorbance with optical fiber spectrometer. The film thickness measuring instrument can accurately measure the thickness of the micron film, and the measurement accuracy can reach $\pm 1.25$ μm. It is proved that the instrument integrates multiple functions, has high measurement accuracy and wide range, and realizes non-contact measurement.

**Keywords:** Multi-functional portable optical measuring instrument, Type Y optical fiber, optical spectrometer

## 1.Introduction

Spectral analysis is the core function of a multifunctional portable optical measuring instrument. Based on this, the multifunctionality of the instrument can achieve spectral detection of various light sources and measure various parameters through spectral analysis, such as solution concentration, pigment content, phase of optical elements, and thickness of optical films.

Spectroscopy can be traced back to the late 17th century when Newton decomposed sunlight into a spectrum of colors using a prism. It has now been widely applied in fields such as chemistry, biochemistry, and environmental monitoring to analyze the composition and structure of samples. Imaging spectrometers are highly efficient quantitative detection instruments that can obtain continuous monochromatic spectral images of the target or scene under observation, providing researchers with spatial and spectral feature details of every point in the target or scene [1-3]. Common types of grating-based imaging spectrometers include the Czerny-Turner imaging spectrometer that uses a plane reflective grating[4-6], the concentric type imaging spectrometer that uses convex and concave gratings (mainly Offner and Dyson imaging spectrometers)[7-9], and the PG or PGP type imaging spectrometer that combines a prism with a transmission grating as the core spectroscopic device[10-11]. Spectroscopic technology is a special discipline that integrates traditional optics with emerging technologies and will continue to gain new vitality with the increasing demands of humanity and the development of technology[12-15].

With the rapid development of science and technology and the increasing demand for measurement accuracy, optical measurement is an important technique in scientific research and is of great significance in promoting scientific studies.[16-19] Traditional optical instruments often have the disadvantages of complex operation, high cost, difficult maintenance, and single function.Against this backdrop, the multifunctional optical measuring instrument, as a detection tool with low cost and simple operation, has the advantages of high measurement accuracy, non-contact and pollution-free, and a wide range of applications.

## 2. Detection principle

The LED (Light Emitting Diode) chip is a semiconductor device that converts electricity into light energy. When a forward voltage is added to both ends of the LED chip, electrons will flow from the cathode to the anode, and holes will flow from the anode to the cathode. As the current flows through the semiconductor, the electrons and holes recombine at the place where they meet (called the active zone), releasing energy and thus generating photons, whose structure is shown in Figure 1A.

The principle of the CCD photoelectric image sensor is described as: the photosensitive element generates an electron hole pair under illumination conditions, and under the action of the internal electric field of the CCD, the electrons and the hole move in the opposite direction, thus generating an electrical signal proportional to the incident light intensity of the incident light at the output of the sensor, FIG. 1B.

A grating is an optical element with a periodic structure, usually composed of a large number of equally spaced parallel cutting lines or grooves, which act to diffract the incident light and thus decompose the light into its constituent wavelengths.

In this instrument, the formula that determines the position of the principal maxima at each order is called the grating equation, which is the basic equation for the design and use of gratings under normal incidence. For the m-th order principal bright fringe of a transmission grating with wavelength λa, the condition is satisfied:

$$dsin\theta = m\lambda_a \qquad\qquad (2-1)$$

The color resolution capability of the grating refers to the ability to resolve two spectral lines with a very small difference in wavelength. Considering two spectral lines with wavelengths $\lambda$ and $\lambda + \Delta\lambda$, if the distance they are separated by dispersion is such that the intensity maximum of one spectral line coincides with the minimum value next to the maximum of the other spectral line, these two spectral lines can just be resolved. At this time, the wavelength difference $\Delta\lambda$ is the smallest wavelength difference that the grating can resolve, and the color resolution capability of the grating is defined as:

$$\Lambda = \frac{\lambda}{\Delta\lambda} = mN \qquad (2-2)$$

This formula indicates that the color resolution capability of the grating is directly proportional to the spectral order m and the number of grating lines N. Hence, the instrument selects a grating with N = 600 lines, which has a good color resolution capability.

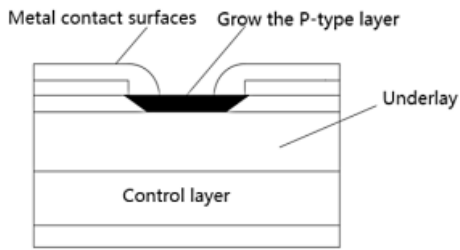

Figure A.the structure of the LED chip

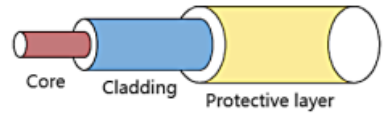

Figure C. Fiber optic construction

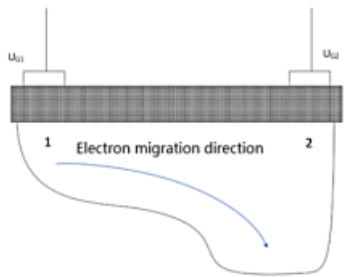

Figure B. schematic diagram of CCD principle

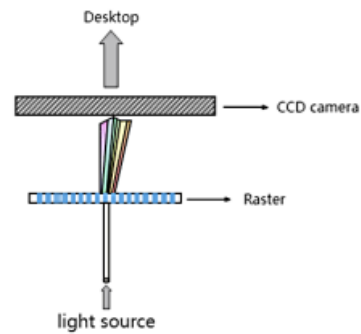

Figure D.The part of optical signal processing

Figure 1

The fibre-optical structure usually consists of three parts: cladding layer, fiber core and protective layer as shown in Figure 1C.

Optical fiber communication system mainly includes transmitter, optical fiber, receiver and detector. In the transmitter, the electrical signal is modulated by the modulator into an optical signal, which is then injected into the optical fiber through the fiber emitter. In the optical fiber, the optical signal is transmitted through the optical fiber to the receiver. In the receiver, the optical signal is converted to an electrical signal through the optical detector, and then restored to the original information through the demodulator.

The instrument is mainly applied to the principle of diffraction grating spectrometer. The structure is shown in Figure 1D.

Working mode is described as: light through the incoming slit, light through diffraction grille, diffraction phenomenon, different wavelength light signal diffusion at different angles, form a series of parallel spectrum band, diffraction after the light signal transfer to the sensor, the sensor will light signal into electrical signal, and recorded through the data acquisition system, finally, through the software to the collected signal data processing and analysis, get the spectral information.

## 3. Instrument design

The enclosure was generated using 3D printing, Is 174 mm long, Is about 174 mm wide, At 50 mm high, Small and portable; The whole appearance is black, Can prevent the influence of external natural light on the internal instruments, See Figure 2A; The material of the shell is PLA (polylactic acid), The material has the advantage that it being fully biodegradable under appropriate conditions, Thus reducing the negative impact on the environment, Very environmentally friendly, Its excellent performance, High strength, high toughness, impact resistance and heat resistance; The shell wall design with small holes, Convenient access to data lines; The shell has a pure black partition plate in the middle, Separate the excitation light source module from the grating diffraction module thoroughly, Avoid the mutual interference of the light signals between the modules. The cover plate is used as the experimental base for the sample placement. The design of the circular texture is conducive to the measurement and both beautiful, as shown in Figure 2C.

Figure 2D and Figure 2E show the interior before and after assembly and fixation, respectively. The whole is divided into light source chamber and spectral chamber, with each baffle in the middle to prevent mutual influence. In the light source room, the LED lamp chip is fixed on the internal wall, and the optical fiber is bent at a certain angle and fixed, so that the optical fiber port is vertically aligned with the LED lamp chip, so that the light source can enter the optical fiber as much as possible. In the spectral room, the CCD camera fixed the base of the spectrometer module, the grating is attached to the lens surface, adjust the position of the optical fiber port, align the slit, so that the light enters the slit as far as possible on the grating, so that the CCD camera receives the complete spectrum, and finally fixed each device.

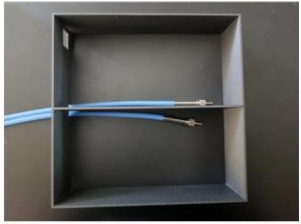
Figure A.Enclosure display

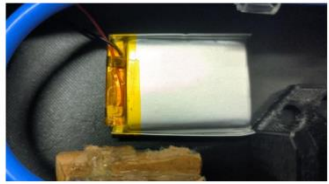
Figure B.Light source battery

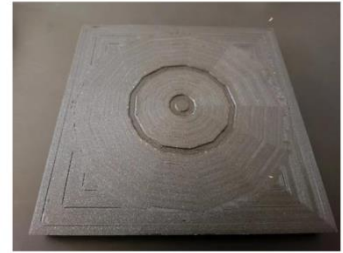
Figure C.Shell lid display

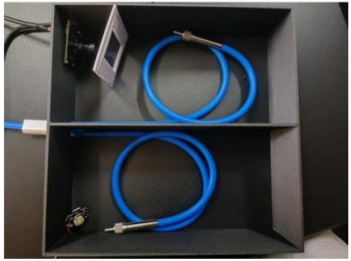
Figure D.Internal parts (not installed)

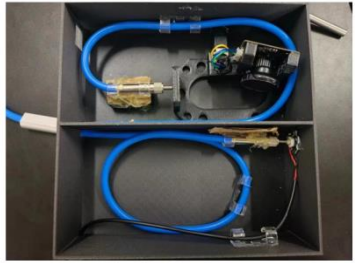
Figure E.Internal parts (installed)

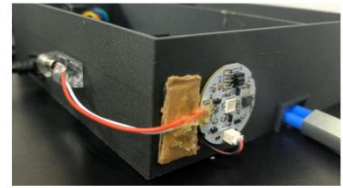
Figure F.Additional multi-color lights

Figure 2

The white LED spectrum is wide, but some experiments have special requirements for the light wavelength, so a 16-color remote controllable light source is added to the instrument to assist the measurement and instrument calibration. Designed to facilitate use, the auxiliary light source with a battery can work for a long time on a single charge. The battery is fixed inside the instrument, as shown in Figure 2B. In order not to affect the internal light source and the spectrometer, the chip is installed on the outer wall of the instrument and connected with DC3.5 charging cable, as shown in Figure 2F.

Figure 3 shows the display picture of the finished product. The cover plate can be used as the experimental base for sample placement; the holder is used for fixing the optical fiber.

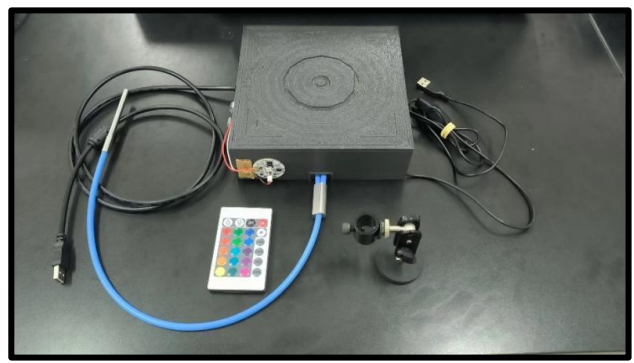

Figure 3

## 4.Test verification

## 4.1 Natural optical spectral detection

In order to verify the spectral performance of the detector, we used the high price scientific research spectrometer and the independently designed detector to conduct spectral detection of natural light, and verified its spectral performance through the comparison of spectral images.

A high-precision spectrometer was used to detect the natural light and obtain the natural spectral image of Figure 4A.

After obtaining the contrast map, we placed the self-designed detector under natural light for detection. The measured natural spectra under different weather, temperature, and environmental conditions were shown in Figure 4B, C and D.

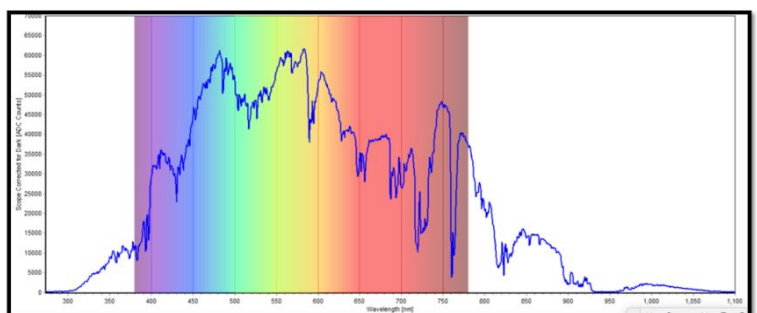

Figure A.Spectra of natural light measured by a high-precision spectrometer

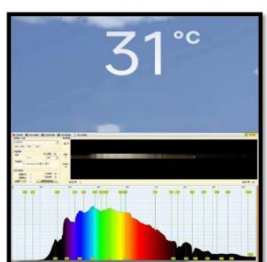 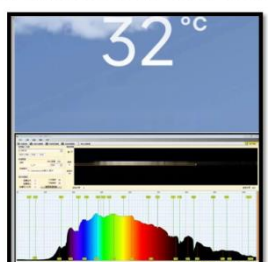 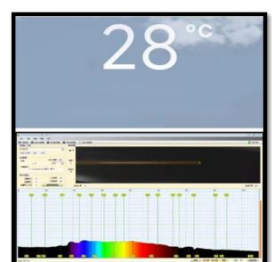

Figure B.July 21, 2024, 10:00, Sunlight Test.　　Figure C.July 22, 2024, 10:00, Sunlight Test.　　Figure D.July 23, 2024, 10:00, Sunlight Test.

Figure 4

Through the comparison of natural light spectrum detection, it can be found that within the visible light range of 380nm-780nm, 380-450nm and 450-495nm, the yellow light at 570-630nm has little influence on the overall trend; the ultraviolet band is roughly the same, and there is slight distortion in the infrared band. The measured data shows that the spectral detection function in the detector is basically the same as that of the high precision spectrometer used in scientific research, which can meet the requirements of the spectral detection performance in the detection. On the other hand, the measured full-band spectra are basically the same on different days but under the same weather conditions, indicating their reliable stability. Based on the above experiments, the observation and analysis of the data show that the observable wavelength range of the spectrometer is about 340-1050nm, and the resolution is 1nm.

## 4.2 Solution concentration monitoring

When A beam of parallel light passes through the uniform solution, the light is absorbed by the solution or through the reflected solution, the absorbance can be calculated by measuring the light intensity, A. According to the Lambert-Beer's law:

$$A = \lg\left(\frac{I_0}{I_1}\right) = Kbc \qquad (4-1)$$

Where K is the mole light absorption coefficient; I0 is the incident light intensity; I1 is the transmitted light intensity; b is the thickness of the absorption layer; and c is the concentration of the light absorbing material.

根 According to the absorbance is proportional to the solution concentration, the software establishes the image and linear function of the absorbance and the solution concentration to obtain the relationship between the relative transmitted light intensity and the solution concentration on the spectral image. The order of samples with different concentrations was disrupted and measurements were performed to verify the correctness of the fitted linear relationship.

First configure potassium permanganate solutions of different known concentrations, Take 0.1g of potassium permanganate, Water, 250ml, Matratio to obtain a concentration of 0.4g / L, That is, the potassium permanganate mother solution of 63.21mol/L, Multiple groups of samples were mixed to obtain the solution concentration of 0.4g / L, 0.3g/L, 0.2g/L, 0.15g/L, 0.1g/L, 0.075g/L, 0.5g/L, At 0.025g / L of potassium permanganate solution, Add the small test tube with the cover into eight groups, Part of the prepared solutions and their self-made spectrometer detection and their detection remote-controlled multicolor light source are shown in Figure 5A.

Second set the incident light and measure the incident spectrum as a control, the eight groups of known concentration solution between the light source and the spectrometer incident fiber, ensure the position, with three-color GRB detection light observation, in the control of the solution concentration of the variable only eight sets of spectral data, after the fixed fiber probe detection concentration, as shown in Figure 5B.

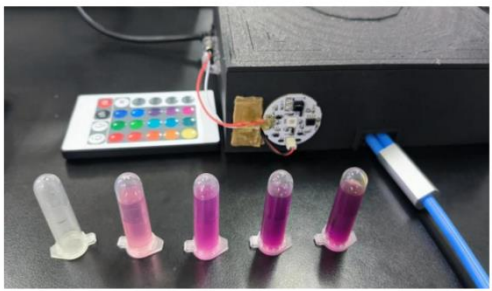 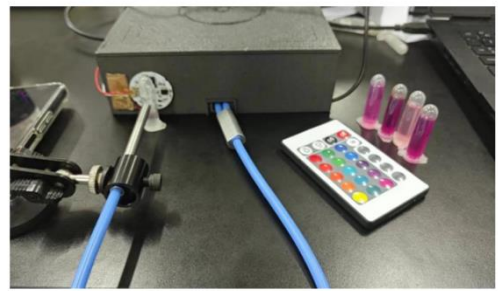

Figure A.Instruments & Samples       Figure B.Conduct experiments

Figure 5

Measurements obtained the spectral data as shown in Figure 5 A-I.

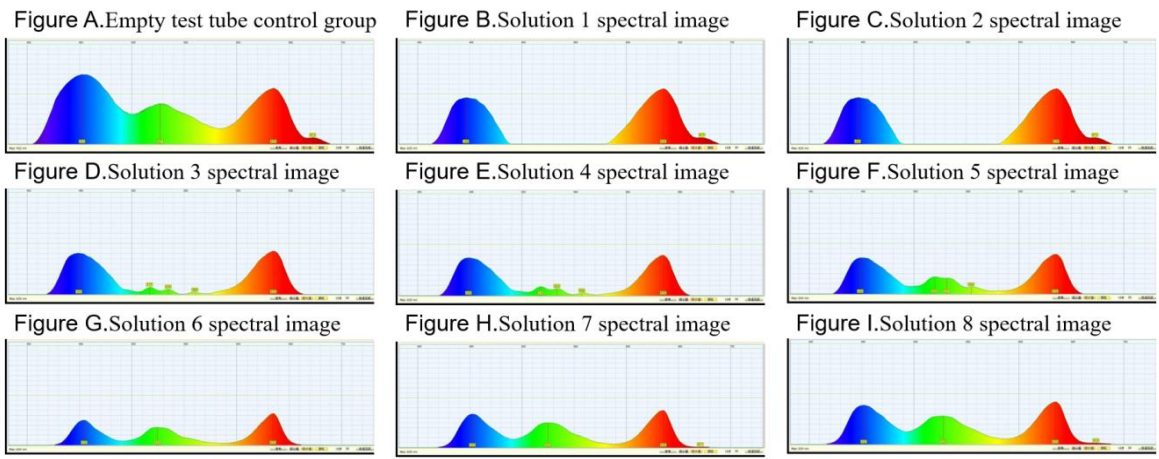

Figure 6

After the spectral data is obtained, compared with the previous incident light spectrum, $KMnO_4$ mainly absorbs the green light, and the absorbance A is obtained, according to the Lambert-Beer law to the known solution concentration, and the curve sample for measuring the potassium permanganate solution is obtained. The experimental results are shown in Table 1 below.

Table 1: The results measured in the experiment

| Group | Solution Concentration（g/L） | Solution Concentration（mol/L） | Relative Incident Light Intensity | Relative Transmitted Light Intensity | absorbance | peak wavelength |
|---|---|---|---|---|---|---|
| solution1 | 0.400 | 63.2136 | 40 | 0 | $\infty$ | None |
| solution2 | 0.300 | 47.4102 | 40 | 0 | $\infty$ | None |
| solution3 | 0.200 | 31.6068 | 40 | 7 | 0.7570 | 517 |
| solution4 | 0.150 | 23.7051 | 40 | 8 | 0.6990 | 518 |
| solution5 | 0.100 | 15.8034 | 40 | 17 | 0.3716 | 520 |
| solution6 | 0.075 | 11.8526 | 40 | 18 | 0.3467 | 524 |
| solution7 | 0.050 | 7.9017 | 40 | 25 | 0.2041 | 525 |
| solution8 | 0.025 | 3.9509 | 40 | 28 | 0.1549 | 528 |

The graph correlation of solution concentration c and absorbance A were again analyzed by Origin software as shown in Figure 6.

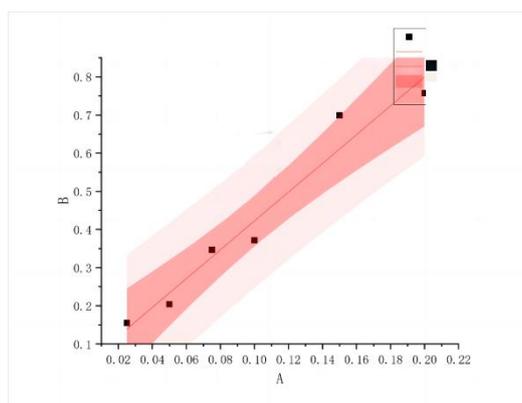

Figure 7

The expression of the linear relationship between concentration and absorbance after fitting is:

$$A = 3.7718c + 0.04504 \qquad (4-2)$$

According to the Lambert-Beer's law:

$$A = \lg\left(\frac{I_0}{I_1}\right) = kbc \qquad (4-3)$$

The relative transmitted light intensity and concentration relationship is deduced:

$$c = 0.2651 \lg\left(\frac{40}{I_1}\right) - 0.012 \qquad (4-4)$$

When the relative incident light is adjusted as the reference (40), the solutions of groups 1 to 8 were remeasured to verify the feasibility of the fitting relationship, and the measurement was randomly sampled and brought into the edited calculation table, and then the concentration data was obtained as shown in Table 2.

Table 2 mapped KMnO$_4$

| Random fetch | Relative transmitted light intensity | Absorbance A | solution concentration（g/L） | solution concentration （mol/L） |
|---|---|---|---|---|
| 1 | 0 | ∞ | None | None |
| 2 | 6 | 0.8239 | 0.2065 | 32.6267 |
| 3 | 7 | 0.7570 | 0.1887 | 29.8223 |
| 4 | 28 | 0.1549 | 0.0291 | 4.6021 |
| 5 | 8 | 0.69898 | 0.1734 | 27.3930 |
| 6 | 0 | ∞ | None | None |
| 7 | 18 | 0.3468 | 0.0800 | 12.6401 |
| 8 | 25 | 0.2041 | 0.0422 | 6.6638 |

For the measured data again, it can be observed that the concentration obtained by the absorbance for different times, which proves that the absorbance measured by the optical fiber spectrometer can be achieved.

In verifying the fitting relationship, the relative transmitted light intensity of some groups is smaller, which may be caused by the distance or Angle of the probe and the light source. For improvement, 3D printing can be used to print the black box of the probe and the light source to fix the sample and reduce the impact of sunlight on the results.

By fitting the relationship function of the absorbance and solution concentration of potassium permanganate solution, it is impossible to achieve the relative transmitted light intensity when the measured concentration is too high, but below the concentration of 0.2g / L, the absorbance can be indirectly obtained by spectral images to measure the concentration of potassium permanganate solution.

## 4.3 Film thickness was measured based on the interference principle

Thin film interference is a manifestation of the interference phenomenon of light on a thin film medium. When a light wave hits the film, it is reflected at the two interfaces due to the different refractive index at the upper and lower interfaces of the film. These reflected light waves will interfere with each other to form a new light wave.

If two beams of coherent light have a constant phase difference $\Delta\varphi$ （$\Delta\varphi=\varphi_2-\varphi_1$） at each specified point in the light field, then the resultant light intensity at a certain point in the encounter space is

$$I = I_1 + I_2 + 2\sqrt{I_1/I_2} \cos\left(\frac{2\pi}{\lambda}\Delta L\right) \qquad (4-5)$$

Where $I_1, I_2$ and $\pi$ can be regarded as constants, then the coherent light intensity is only related to the optical path difference $\Delta L$ and the wavelength $\lambda$.

The formula for calculating the thickness of the film is derived from the relevant formulas:

$$d = \frac{\lambda_1\lambda_2}{2n(\lambda_1 - \lambda_2)} \qquad (4-6)$$

The sample prepared for the test is a PET protective film of unknown thickness on the surface of a building materials.

Place the instrument on the flat table top, lay the sample to be tested on the measurement base, fix the optical fiber probe, and open the LED white light source. As shown in Figure 7B.

Debug the integration time, sampling location and other relevant data on the computer, and save the spectral pictures in Figure 7C.

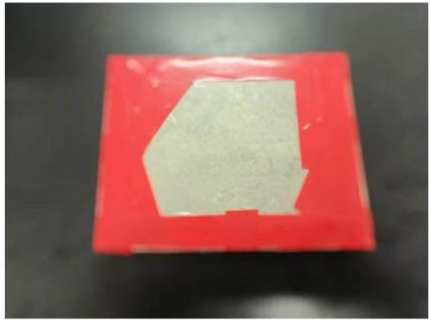 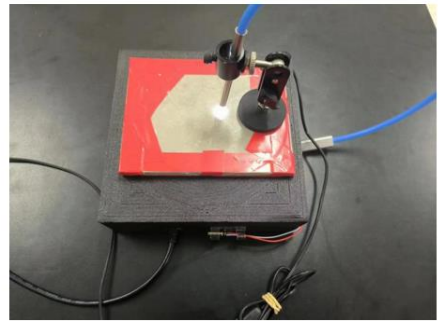

Figure A.Specimen films          Figure B.Take measurements

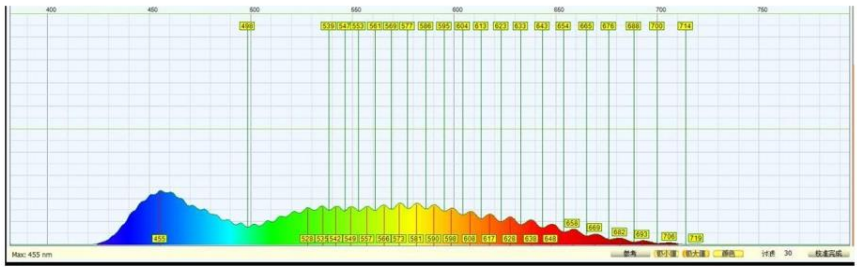

Figure C.Spectral picture

Figure 8

The process of sampling and calculation of the wave crest is to sample the peaks with an even number from the numbers 1-10, that is, to collect the peaks of 2, 4, 6, 8, 10 and other numbers respectively, and sample and calculate the peaks with each even number three times in different bands. The results are plotted as shown in the table.

Table 3 wave crest sampling data

| Peak number / Sampling time | 2 | 4 | 6 | 8 |
|---|---|---|---|---|
| 1 | 11.20μm | 11.24μm | 11.99μm | 12.00μm |
| 2 | 12.18μm | 11.88μm | 11.39μm | 11.62μm |
| 3 | 11.78μm | 11.54μm | 11.46μm | 11.80μm |

Calculate the mean and standard deviation of film thickness, and the relative uncertainty of measurement results:

Table 4 Analysis of result data

| Peak number | 2 | 4 | 6 | 8 | 10 |
|---|---|---|---|---|---|
| Mean value （μm） | 11.72 | 11.55 | 11.61 | 11.80 | 11.69 |
| Standard deviation （μm） | 0.493 | 0.319 | 0.464 | 0.165 | 0.117 |
| Relative uncertainty （μm） | 2.4% | 1.5% | 2.3% | 0.80% | 0.58% |

Analysis of results: The thickness of the PET film of the sample is in the range of 10-20μm, and the thickness of the PET film is in the range of 11.50μm to 11.80μm, regardless of the influence of the external environment and the error allowed. The Class A uncertainty and standard deviation calculated according to the different number of peak values can be seen that increasing the group is conducive to reducing the uncertainty of the measurement results. Reflecting a significant increase in the reliability of the measurement. It can significantly improve the accuracy and reliability of experimental measurement data, especially in applications requiring high precision measurements.

## 5. Conclusion

In this paper, a multi-functional portable optical measuring instrument is designed based on "Y" type optical fiber. Compared with the shortcomings of traditional optical measuring instrument, such as single function, large volume and complex operation, the multi-functional portable optical measuring instrument is a portable, simple operation and low cost detection tool, which has the characteristics of multi-functionality, high measurement accuracy, non-contact and pollution-free, wide application range and so on.

Based on the principle of grating diffraction, the principle of optical fiber transmission, the principle of light interference and so on, the multifunctional portable optical measuring instrument has designed and assembled a visual spectrometer for optical measurement based on the CCD photoelectric image sensor. It has improved and designed a "Y"-shaped optical fiber optical path for light signal transmission applicable to multifunctional measurement. Combined with remotely controllable multi-color LED lights, a multifunctional optical measurement system has been built.

The main functions of the measuring instrument are spectral measurement and film thickness measurement. It adopts the spectral analysis technology combining CCD and grating. The slit grating is used to disperse the received light to form a spectrum, and the CCD camera receives it and finally converts it into an electrical signal. As for film thickness measurement, it is based on the analysis and calculation of spectral data. The bands of interference and reflection fringes generated by light on the film to be measured are analyzed by the spectrometer, and the thickness of the film to be measured is calculated by the equal thickness interference formula.

Through experiments, the instrument has realized spectral analysis, solution concentration monitoring and film thickness measurement. The experimental results show that the wavelength range of the spectrometer is large and the resolution is high. The solution concentration can be obtained by measuring absorbance with optical fiber spectrometer. The film thickness measuring instrument can accurately measure the thickness of micron-level films, and its measurement accuracy can reach $\pm 1.25$ μm. It has been verified that this instrument integrates multiple functions into one, has high measurement accuracy and a wide measurement range, and realizes non-contact measurement. Moreover, on the basis of simple operation, the instrument makes the phenomena more intuitive and obvious. It directly displays the film thickness data of the object to be measured through the operation formula, which greatly reduces errors and mistakes. Therefore, this instrument has great advantages in scientific research and production, and the multi-function also makes it highly extensible.

**Acknowledgments.** National Natural Science Foundation of China (62305262). Shaanxi Fundamental Science Research Project for Mathematics and Physics (22JSQ026). Shaanxi Province College Student Innovation and Entrepreneurship Training Program Project (202410703052).

**Competing interests.** The authors declare that they have no competing interests.

**Data availability.** Data underlying the results presented in this paper are not publicly available at this time but may be obtained from the authors upon reasonable request.

# Reference

[1] Mouroulis P,Green R O.Review of high fidelity imaging spectrometer design for remote sensing [J].Optical Engineering,2018,57(4):040901.

[2] Lumb D H,Bautz M W,Burrows D N,et al.Recent developments for the AXAF CCD imaging spectrometer [C]//EUV,X-Ray, and Gamma-Ray Instrumentation for Astronomy IV,International Society for Optics and Photonics, 2006: 265-271.

[3] Kudenov M W,Dereniak E L.Compact real-time birefringent imaging spectrometer[J].Optics Express,2012,20(16):17973 17986.

[4] M.V.R.K.Murty.Use of convergent and divergentillumination with plane gratings[J]. J.Opt.Soc.Am.,1962,52(7):768-773

[5] Aathur B. Shafer. Optimization of the Czerny-Turner Spectrometer [J]. J.Opt.Soc.Am., 1963,54(7):

[6] Murphy L.Dalton,Jr.Astigmatism compensation in the Czerny-Turner spectrometer [J].Appl.Opt.,1966,5(7):1121-1123

[7] Pavlycheva N K.Sov.J.Opt.Technol.1979,46(7):394.

[8] Hutley M C.Diffraction Gratings.Academic Press,1980.

[9] ZHOU Qian,ZENG L-i jiang,LI L-i feng.Spectroscopy and Spectral Analysis,2008,28(7):1673.

[10] Li Q C,Jiang Y J.Principle of spectroscopic instruments[M].Beijing:Machinery Industry Press, 1989:3-1 5.

[11] Ju H,Wu Y H.The development status of micro-spectrometer[J].Optics And Precision Engineering,2001,9(4):372-376.

[12] Korablev O,Montmessin F,Trokhimovsky A,et al.Compact echelle spectrometer for oc cultation sounding of the Martian atmosphere:design and performance [J].Applied Optics,2013,52(5):1054-65.

[13] Tousey R,Purcell J D,Garrett D L,et al.An echelle spectrograph for middle ultraviolet solar spectroscopy from rockets[J].Applied Optics,1967,6(3):365.

[14] Guo H,Xiao G,Mrad N,et al.Echelle Diffractive Grating Based Wavelength Interrog ator for Potential Aerospace Applications[J].Journal of Lightwave Technology,2013,3 1(13):2099-105.

[15] Kane R,Siegmund O H,Beasley M,et al.The opto-mechanical design of the Colorado High-resolution Echelle Stellar Spectrograph(CHESS)[J].Proceedings of SPIE-The Inter national Society for Optical Engineering,201 1,8 145:8 1450P-P-8.

[16] Liu X L,Liu X,Li W,et al.Optical System Design of Space-Based Filament LiDAR Spectr ometer[J].Chinese Journal of Lasers,2023.50(7),0708012.

[17] Liu Y Q,Jiang C,Liu Z Y,et al.Long-Period Fiber Gratings[J].Laser&Optoelectronics Pr ogress,2023.60(9),0900001.

[18] Sun Y C,Huang C,Xia G,et al.Accurate wavelength calibration method for compact CCD spectrometer[J].Journal ofthe Optical Society ofAmerica A:2017.34(4):498-505.

[19] Ballester P,Rosa R M P Ballester,Rosa M R.Modeling echelle spectrographs[J].Astronomy and Astrophysics:1997,126:563-571.